\documentclass[12pt]{article}
\usepackage{amsthm,amssymb,amsmath}
\usepackage[english]{babel}
 1 
1

\newcommand{\C}{\ensuremath{\mathbb{C}}}

\newcommand{\Z}{\ensuremath{\mathbb{Z}}}

\newcommand{\p}{{\boldsymbol{p}}}
\newcommand{\Q}{{\boldsymbol{Q}}}

\newcommand{\bu}{{{\bf u}}}

\newcommand{\ba}{{{\bf a}}}
\newcommand{\bR}{{\boldsymbol{R}}}

\newtheorem{teh}{Theorem}

\newcommand{\be}{\begin{equation}}
\newcommand{\ee}{\end{equation}}
\begin{document}

\title{\sc Integrable Quasiclassical Deformations of Cubic
Curves.
\thanks{Partially supported by  DGCYT
project BFM2002-01607 and by the grant COFIN 2004 "Sintesi" }}
\author{Y. Kodama $^{1}$\thanks{Partially
supported by NSF grant DMS0404931}, B. Konopelchenko $^{2}$,\\
 L. Mart\'{\i}nez Alonso$^{3}$ and E. Medina$^{4}$
\\
\emph{ $^1$ Department of Mathematics, Ohio State University}
\\ {\em Columbus, OH 43210, USA}
\\
\emph{ $^2$ Dipartimento di Fisica, Universit\'a di Lecce and
Sezione INFN}
\\ {\emph 73100 Lecce, Italy}\\
\emph{$^3$ Departamento de F\'{\i}sica Te\'{o}rica II, Universidad
Complutense}\\ \emph{E28040 Madrid, Spain}\\
\emph{$^4$ Departamento de Matem\'aticas, Universidad de
C\'adiz}\\ \emph{E11510 Puerto Real, C\'adiz,Spain} }
\date{} \maketitle
\begin{abstract}
A general scheme for determining and studying hydrodynamic type systems
describing  integrable deformations of algebraic curves is applied to cubic
curves. Lagrange resolvents of the theory of
cubic equations are used to derive and characterize these
deformations.
\end{abstract}

\vspace*{.5cm}

\begin{center}\begin{minipage}{12cm}
\emph{Key words:}  Algebraic curves. Integrable systems. Lagrange
resolvents.

\emph{PACS number:} 02.30.Ik.
\end{minipage}
\end{center}
\newpage

\section{Introduction}

The theory of algebraic curves is a fundamental ingredient  in the
analysis of integrable nonlinear differential equations as it is
shown, for example, by its relevance in the description of the
finite-gap solutions or the formulation of the Whitham averaging
method \cite{1}-\cite{8}. A particularly interesting problem is
characterizing and classifying  integrable deformations of
algebraic curves. In \cite{6}-\cite{7} Krichever formulated a
general theory of dispersionless hierarchies of integrable models
arising in the Whitham averaging method. It turns out that the
algebraic orbits of the genus-zero Whitham equations determine
infinite families of integrable deformations of a particular class
of algebraic curves. A different approach for determining
integrable deformations of general algebraic curves $\mathcal{C}$
defined by monic polynomial equations
\begin{equation}\label{1}
\mathcal{C}:\quad F(p,k):=p^N-\sum_{n=1}^{N}u_n(k)p^{N-n}=0,\quad u_n\in \C[k],
\end{equation}
was proposed in \cite{9}-\cite{10}. It applies for finding
deformations $\mathcal{C}(x,t)$ of $\mathcal{C}$ with the
deformation parameters $(x,t)$, such that the multiple-valued
function $\p=\p(k)$ determined by \eqref{1} obeys an equation of
the form of conservation laws
\begin{equation}\label{2}
\partial_t \p=\partial_x \Q,
\end{equation}
where the flux $\Q$ is given by an element from ${\mathbb C}[k,p]/{\mathcal{C}}$,
\[
\Q=\sum_{r=1}^{N}a_r(k,x,t)\p^{r-1},\quad a_r\in\C[k].
\]
Starting with \eqref{2}, changing to the dynamical variables $u_n$
and using Lenard-type relations (see \cite{10}) one gets a scheme
for finding consistent deformations of \eqref{1}. One should also
note that \eqref{2} provides an infinite number of conservation
laws, when one expands $\p$ and $\Q$ in Laurent series in $z$ with
$k=z^r$ for some $r$. In this sense, we say that equation
\eqref{2} is integrable.

 Our strategy can be applied to the \emph{generic
case} where  the coefficients (\emph{potentials})  $u_n$ of
\eqref{1} are general polynomials in $k$
\[
u_n(k)=\sum_{i=0}^{d_n} u_{n,i}k^i,
\]
with  all the coefficients $u_{n,i}$ being considered as
independent dynamical variables, i.e. $u_{n,i}=u_{n,i}(x,t)$.
However, with appropriate modifications, the scheme  can be also
applied to cases in which constraints on the potentials are
imposed. A complete description of these deformations for the
generic case of hyperelliptic curves ($N=2$) was given in
\cite{10}.

The present paper is devoted to the deformations of cubic curves
($N=3$)
\begin{equation}\label{3}
p^3-w\,p^2-v\,p-u=0, \quad u,v,w \in {\mathbb C}[k]\,,
\end{equation}
and it considers not only  the generic case but also the important
constrained case $w\equiv 0$.
Although some of the curves may be conformally equivalent
(with for example the dispersionless Miura transformation), we will
not discuss the classification problem under this equivalence in this paper
(we will discuss the details of the problem elsewhere).
In section 2 a general approach to
construct integrable deformations of algebraic curves is reported
briefly. Section 3 is devoted to the analysis of the cubic case
\eqref{3}. We emphasize the role of Lagrange resolvents, describe
the Hamiltonian structure of integrable deformations and  present
several illustrative examples including Whitham type deformations.

\section{Schemes of deformations of algebraic curves}

In order to describe  deformations of the curve $\mathcal{C}$
defined by \eqref{1}, one may use the potentials $u_n$, as well as
the $N$ branches $p_i=p_i(k)$ ($i=1,\ldots,N$)  of the
multiple-valued function function $\p=\p(k)$ satisfying
\begin{equation}\label{4}
F(p,k)=\prod_{i=1}^{N} (p-p_i(k)).
\end{equation}

  The potentials
 can be expressed as elementary symmetric polynomials $s_n$
\cite{11}-\cite{13} of the branches $p_i$
\begin{equation}\label{5}
u_n=(-1)^{n-1}s_n(p_1,p_2,\ldots)= (-1)^{n-1}\sum_{1\leq i_1<\ldots<i_n\leq
N}p_{i_1}\cdots p_{i_n}.
\end{equation}
However, notice that, according to the famous Abel theorem
\cite{11}, for $N>4$ the branches $p_i$ of the generic equation
\eqref{1} cannot be written in terms of the potentials $u_n$ by
means of rational operations and radicals.

There is an important result concerning the branches $p_i$ which is useful in our
analysis. Let
$\C((\lambda))$ denote the field of Laurent series in $\lambda$ with
at most a finite number of terms with positive powers
\[
\sum_{n=-\infty}^N c_n \lambda^n,\quad N\in \Z.
\]
Then we have \cite{14,15} :

\begin{teh}({\bf Newton Theorem})

There exists a positive integer $l$ such that the $N$ branches
\begin{equation}\label{6}
p_i(z):=\Big(p_i(k)\Big)\Big |_{k=z^l},
\end{equation}
are elements of  $\C((z))$. Furthermore, if $F(p,k)$ is irreducible as a polynomial over the
field $\C((k))$ then $l_0=N$ is the least permissible $l$ and the
branches $p_i(z)$ can be labelled so that
\[
p_i(z)=p_N(\epsilon^i z),\quad \epsilon:=\exp \frac{2\pi i}{N}.
\]
\end{teh}

\noindent {\bf Notation convention} \emph{
 Henceforth, given an algebraic curve $\mathcal{C}$ we
will denote by $z$ the variable associated with the least positive
integer $l_0$  for which the substitution $k=z^{l_0}$ implies
$p_i\in\C((z)),\, \forall i$. The number $l_0$ will be referred to
as the Newton exponent of the curve }.

\vspace{0.3truecm}

 For the generic case the method proposed in \cite{10}  may be
summarized as follows : Given an algebraic curve \eqref{1}, we define
an evolution equation of the form
\begin{equation}\label{7}
\partial_t \bu = J_0\Big(T\,\nabla_{\bu} R\Big)_{+},
\quad R(z,\p)=\sum_i f_i(z)\,p_i,
\end{equation}
where $(\,\cdot\,)_+$  indicates the
part of non-negative  powers  of a Laurent series in $k$
and
\[
f_i\in \C((z)),\quad \nabla_{{\bf u}}R:=\Big(\frac{\partial
R}{\partial u_1}\ldots \frac{\partial R}{\partial
u_N}\Big)^{\top},
\]
\begin{equation}\label{8}
J_0:= T^{\top} V^{\top}\partial_x V,
\end{equation}
\begin{equation}\label{9}
T:=\left(
\begin{array}{cccc}
1&-u_{1}&\cdots&-u_{N-1}\\
0&1&\cdots&-u_{N-2}\\
\multicolumn{4}{c}{\dotfill}\\
0&\multicolumn{2}{c}{\dotfill}&1
\end{array}
\right),\quad  V:=\left(
\begin{array}{cccc}
1&p_1&\cdots&p_1^{N-1}\\
\multicolumn{4}{c}{\dotfill}\\
\multicolumn{4}{c}{\dotfill}\\
1&p_N&\cdots&p_N^{N-1}
\end{array}
\right).
\end{equation}
Let $d_{nm}$ and $d_n$ be the degrees of the matrix elements $(J_0)_{nm}$
and the potentials $u_n$ as polynomials in $k$, respectively.
Then \eqref{7} defines a  deformation of the curve, if  $d_{nm}$ and $d_n$
satisfy the consistency conditions
\begin{equation}\label{10}
\max\{d_{nm},m=1,2,3\}\,\leq\, d_n+1,\quad n=1,2,3,
\end{equation}
and  the components of  $\nabla_{\bu}R$ are in
$\C((k))$ with $k=z^{l_0}$.

Equivalently, in terms of branches
\[
{\bf p}:=\left(p_1,\ldots, p_{N} \right)^{\top},
\]
the system \eqref{7} can be written as
\begin{equation}\label{11}
\partial_t\,{\bf p}=\partial_x(V\,{\bf r}_+),
\end{equation}
where
\begin{equation}\label{12}
{\bf r}:=T\,\nabla_{\bu}R(z,\p)=V^{-1}{\bf f}(z),
\end{equation}
with ${\bf f}(z):= \left(f_1(z),\ldots,
f_{N}(z) \right)^{\top}$.
Notice that ${\bf r}$ is a solution of the Lenard relation
\begin{equation}\label{13}
J_0{\bf r}=0.
\end{equation}

Although there is not a general procedure for analyzing
constrained
 cases, one may try a similar strategy. Firstly,
we start from the equation for branches  \eqref{2} and then, by
expressing the potentials in terms of the independent branches only,
we look for a formulation of the flows as
\begin{equation}\label{14}
\partial_t \bu = J_0{\bf a},\quad {\bf a}:=(a_1,\ldots, a_N)^{\top},
\end{equation}
for a certain operator $J_0$. Finally, we use solutions ${\bf r}$
of Lenard relations \eqref{13} and set ${\bf a}={\bf r}_+$.

\vspace{0.3truecm}

Another scheme for defining integrable deformations of algebraic
curves of genus $zero$ (i.e. rational curve) is implicit in the
theory of integrable systems of dispersionless type developed in
\cite{7,8}, which we refer to as the Whitham deformations.
It concerns with algebraic curves characterized
by equations of the form
\begin{equation}\label{15}
k=p^{N}+v_{N-2}p^{N-2}+\cdots+v_0+\sum_{r=1}^{M}
\sum_{i=1}^{n_r}\frac{v_{r,i}}{(p-w_{r})^i},
\end{equation}
where $v_n,\,v_{r,i},\,w_{r}$ are $k$-independent coefficients.
These curves arise in the theory of algebraic orbits of the
genus-zero Whitham hierarchy \cite{7,8}, where the function $k$
represents the Landau-Ginzburg potential of the associated
topological field theory. We may rewrite the equation of the curve
\eqref{15} in the polynomial form \eqref{1} with potentials $u_n$
of degrees $d_n\leq 1$ and satisfying a certain system of
constraints.

To describe the deformations of \eqref{15} determined by Whitham
flows we introduce local coordinates $\{z_0,z_1,\ldots,z_{M}\}$ of
the extended $p$-plane at the punctures $\{w_0:=\infty,w_1,\ldots,
w_{M}\}$ such that
\begin{equation}\label{16}
k=z_0^{N}=z_1^{n_1}=\ldots=z_{n_{M}}^{n_{M}}.
\end{equation}
It is clear that there are $N$ branches of $\p$ which have
expansions in powers of $k^{1/N}$ and that, for each puncture
$w_r,\;(r=1,\ldots, M) $, there are $n_r$ branches of $\p$ having
expansions in powers of $k^{1/n_r}$. Therefore, the Newton
exponent $l_0$ is given by the least common multiple of the set of
integers $\{N,n_1,\ldots,n_{M}\}$. Furthermore, it is clear that
only in the absence of finite punctures ($M=0$) the curve
\eqref{15} is irreducible over $\C((k))$.

At each puncture in $\{\infty,w_1,\ldots, w_{M}\}$, there is an
infinite family of Whitham deformations of \eqref{15}. They can be
expressed by equations of the form (see \cite{7,8})
\begin{equation}\label{17}
\partial_t \p=\partial_x \Q_{\alpha,n},
\end{equation}
where
\begin{equation}\label{18}
\left\{\begin{array}{llll}
\nonumber \Q_{\alpha,n}=&(z_{\alpha}^n)_{\oplus}(\p),\quad \alpha=0,1,\ldots,M,\; n\geq 1\\\\
\nonumber \Q_{r,0}=&\ln (\p-w_r),\quad r=1,\ldots,M.
\end{array}\right.
\end{equation}
Here $(z_{\alpha}^n)_{\oplus}$ stands for the singular part of
$z_{\alpha}^n(p)$ at the puncture $w_{\alpha}$, with
$(z_{r}^n)_{\oplus}(\infty)=0$ for $1\leq r\leq M$. There exist
also commuting flows for the negative $n$ in (\ref{17}) with
logarithmic terms which correspond to the descendant flows of $Q_{r,0}$
(see \cite{8} for the details).

In the absence of finite punctures ($M=0$),  Whitham deformations
become the dispersionless Gelfand-Dikii flows. They can be
described by our scheme \cite{10} as the reductions $u_1\equiv
0,\, u_{N}=k-v_0$ of the generic case  corresponding to
$d_n=\delta_{Nn}$. However, for $M\geq 1$ it can be seen that, in
general, Whitham deformations of \eqref{15} are not reductions of
the flows \eqref{7} provided by our method. Some examples of this
situation for cubic curves are shown below.

\section{Deformations of cubic curves}

For our subsequent analysis we introduce a basic tool of the
theory of third order polynomial equations \cite{11}: the so
called \emph{Lagrange resolvents}, defined by
\begin{equation}\label{19}
\mathcal{L}_i:=\sum_{j=1}^3 (\epsilon^i)^j\,p_j,\quad
i=1,2,3,\quad \epsilon:=e^{\frac{2\pi\,i}{3}},
\end{equation}
or, equivalently,
\begin{equation}\label{20}
\left\{\begin{array}{llll}
\nonumber\mathcal{L}_1:&=\epsilon\, p_1+\epsilon^2\,p_2+p_3,\\
\mathcal{L}_2:&=\epsilon^2\,p_1+\epsilon\,p_2+p_3,\\
\nonumber \mathcal{L}_3:&=p_1+p_2+p_3,
\end{array}\right.\end{equation}
They can be expressed in terms of the potentials $\bu=(w,v,u)^\top$ by
using the identities
\begin{equation}\label{21}
\left\{\begin{array}{lll}
\nonumber \mathcal{L}_1\cdot \mathcal{L}_2=3\,v+w^2,\quad
\mathcal{L}_3=w,\\\\
\nonumber \mathcal{L}_1^3+ \mathcal{L}_2^3= 27u+9vw+2\,w^3,
\end{array}\right.\end{equation}
which lead to
\begin{equation}\label{22}
\left\{\begin{array}{lll}
\nonumber 2\,\mathcal{L}_1^3= 27u+9vw+2\,w^3&+\sqrt{(27u+9\,vw+2\,w^3)^2-4(3v+w^2)^3},\\\\
\nonumber 2\,\mathcal{L}_2^3=
27u+9vw+2\,w^3&-\sqrt{(27u+9\,vw+2\,w^3)^2-4(3v+w^2)^3}.
\end{array}\right.\end{equation}
The fundamental advantage of Lagrange resolvents is that they
provide explicit expressions of  the branches $p_i$ in terms of the potentials
according to Cardano formulas
\begin{equation}\label{23}
3\,p_i=\sum_{j=1}^3 (\epsilon^{-i})^j\,\mathcal{L}_j, \quad i=1,2,3,
\end{equation}
or, equivalently,
\begin{equation}\label{24}
\left\{\begin{array}{lll}
\nonumber 3\,p_1&=\epsilon^2\,\mathcal{L}_1+\epsilon\,\mathcal{L}_2+\mathcal{L}_3,\\
3\,p_2&=\epsilon\,\mathcal{L}_1+\epsilon^2\,\mathcal{L}_2+\mathcal{L}_3,\\
\nonumber 3\,p_3&=\mathcal{L}_1+\mathcal{L}_2+\mathcal{L}_3.
\end{array}\right.\end{equation}
As we will prove below, the Lagrange resolvents are essential to
determine consistent deformations of cubic equations.

\subsection{Generic case}

As it was found in \cite{10},  for $N=3$ the operator $J_0$ reads
\begin{equation}\label{25}
J_0 =
\left(
\begin{array}{ccc}
3\partial_x & w\partial_x+w_x&(2v+w^2)\partial_x+(2v+w^2)_x \\\\
-2w\,\partial_x& 2v\,\partial_x+v_x&(3u+vw)\partial_x+2u_x
+2vw_x\\\\
-v\partial_x&3u\partial_x+u_x&uw\partial_x+2u\,w_x
\end{array}
\right)
\end{equation}
Thus if $d_1,d_2$ and $d_3$ are the degrees in $k$ of the potential functions $w,\,v$ and $u$, respectively, the consistency conditions
 \eqref{10} are
\begin{equation}\label{26}
\left\{\begin{array}{lll}
\nonumber &d_1\leq 1,\quad d_2\leq d_1+1,\\\\
\nonumber &d_3\leq d_2+1,\quad d_2\leq d_3+1,
\end{array}\right.\end{equation}
which lead to the following twelve nontrivial choices for
$(d_1,d_2,d_3)$
\begin{align}\label{27}
\nonumber &(0,0,1),\,(0,1,0),\, (0,1,1),\, (0,1,2), \\
&(1,0,0),\, (1,0,1),\,(1,1,0),\,(1,1,1),\\
\nonumber &(1,1,2),\,(1,2,1),\,(1,2,2),\, (1,2,3).
\end{align}
%Note that if any of the consistency conditions in \eqref{10} is
%satisfied as a strict inequality, then  one has that some of the
%coefficients of the potentials does not depend on time. More
%precisely, if we write
%$$\everymath{\displaystyle}\begin{array}{c}
%w(k,x,t)=\sum_{j=0}^{d_1}w_j(x,t)k^j,\quad
%v(k,x,t)=\sum_{j=0}^{d_2}v_j(x,t)k^j,\\  \\
%u(k,x,t)=\sum_{j=0}^{d_3}u_j(x,t)k^j,\end{array}$$
%we find that
%
%\begin{eqnarray}
%\label{case1}
%\mbox{In the case} &(0,0,1): & w_{0\,t}=0,\quad
%u_{1\,t}=0.\\ \nonumber \\
%\label{case2}
%\mbox{In the case} & (0,1,2): &  u_{2\,t}=0.\\  \nonumber \\
%\label{case3}
%\mbox{In the case} & (0,1,0): & v_{1\,t}=0.\\ \nonumber  \\
%\label{case4}
%\mbox{In the case} & (0,1,1): & v_{1\,t}=0,\quad
%u_{1\,t}=0.\end{eqnarray}

By using \eqref{21}-\eqref{24} it is straightforward to determine
the Newton exponent $l_0$ for each of the cases \eqref{27}. Thus
one finds three categories

\vspace{0.5truecm}

 \hfil\vbox{\hbox{\vbox{\offinterlineskip
 \halign{&\vrule#&\strut\quad#\hfil\quad&\vrule#&
 \quad#\hfil\quad&\vrule#&
 \quad#\hfil\quad\cr
 % '\quad#\hfil\quad' in the lines above
 % gives left aligned entries. Substitute
 % `\quad\hfil#\hfil\quad' or '\quad\hfil#\quad'
 % for centered or right aligned.
 \noalign{\hrule}
 height2pt&\omit&&\omit&&\omit&&\omit&\cr
 &$l_0$&&\hspace{1.3truecm}3&&\hspace{1.3truecm}2&&\hspace{1.3truecm}1&\cr
 height2pt&\omit&&\omit&&\omit&&\omit&\cr
 \noalign{\hrule}
 height2pt&\omit&&\omit&&\omit&&\omit&\cr
   &  && (0,0,1),\,(0,1,2) && (0,1,0),\,(0,1,1) && (1,0,1),\,(1,1,0) &\cr
   & &&  && (1,0,0),\,(1,1,2) && (1,1,1),\,(1,2,1) &\cr
   &  &&  &&  && (1,2,2),\,(1,2,3) &\cr
 height2pt&\omit&&\omit&&\omit&&\omit&\cr
 \noalign{\hrule}}}}}\hfill

\medskip

Only the  cases with $l_0=3$ correspond to irreducible curves over the field
$\C((k))$. We also note here that our deformations for the trigonal curves
\eqref{1} in the generic case allow one to have
only the curves with genus less than or equal to one (the details will be
discussed elsewhere).

Once the Newton exponent $l_0$ is known, in order to derive the
associated hierarchy of integrable deformations according to our
scheme, two steps are still required:
\begin{enumerate}
\item To determine the functions $R(z,\p)=\sum_i f_i(z)\,p_i$
such that the components of $\nabla_{\bu}R$ are in $\C((k))$
with $k=z^{l_0}$.
\item To find the explicit form of the gradients $\nabla_{\bu}R$ in
terms of the potentials.
\end{enumerate}
Both problems admit a convenient treatment in terms of Lagrange
resolvents. Thus by introducing the following element $\sigma_0$
of the Galois group of the curve
\begin{equation}\label{28}
\sigma_0(p_i)(z):=p_i(\epsilon_0\,z),\quad
\epsilon_0:=e^{\frac{2\pi i}{l_0}},
\end{equation}
we see that our first  problem can be fixed by determining
functions $R$ invariants under $\sigma_0$  i.e.
$R(\epsilon_0\,z,\sigma_0\,\p)=R(z,\p)$. In this way, we have the following
forms of $R$:

\vspace{0.3truecm}
For the case ${l_0=3}$,
the element $\sigma_0$ is given by the permutation
\begin{equation}\label{29}
\sigma_0=\left(
\begin{array}{ccc}
p_1 & p_2&p_3 \\
p_2 & p_3&p_1
\end{array}
\right),
\end{equation}
or, in terms of Lagrange resolvents,
\begin{equation}\label{30}
\sigma_0=\left(
\begin{array}{ccc}
\mathcal{L}_1 & \mathcal{L}_2&\mathcal{L}_3 \\
\epsilon^2 \mathcal{L}_1 & \epsilon \mathcal{L}_2& \mathcal{L}_3
\end{array}
\right).
\end{equation}
Thus we get the invariant functions
\begin{equation}\label{31}
R=zf_1(z^3)\,\mathcal{L}_1+z^2f_2(z^3)\, \mathcal{L}_2+f_3(z^3)\,
\mathcal{L}_3,
\end{equation}
with $f_i(z^3)$ being arbitrary functions in $\C((z^3))$.

\vspace{0.3truecm}
For the case ${l_0=2}$,
$\sigma_0^2$ is the identity
permutation, so that under the action of $\sigma_0$ two branches
are interchanged  while the other remains invariant. If we label the
branches in such a way that
\begin{equation}\label{32}
\sigma_0=\left(
\begin{array}{ccc}
p_1 & p_2&p_3 \\
p_2 & p_1&p_3
\end{array}
\right),
\end{equation}
then
\begin{equation}\label{33}
\sigma_0=\left(
\begin{array}{ccc}
\mathcal{L}_1 & \mathcal{L}_2&\mathcal{L}_3 \\
\mathcal{L}_2 & \mathcal{L}_1& \mathcal{L}_3
\end{array}
\right),
\end{equation}
and we obtain the invariant functions
\begin{equation}\label{34}
R=f_1(z^2)\Big( \mathcal{L}_1+\mathcal{L}_2\Big)+zf_2(z^2)\Big(
\mathcal{L}_1- \mathcal{L}_2\Big)+f_3(z^2)\, \mathcal{L}_3,
\end{equation}
where $f_i(z^2)$ are arbitrary functions in $\C((z^2))$.

\vspace{0.3truecm}
For the case ${l_0=1}$, we have $z=k$ and $\sigma_0$ is the identity,
so that any function $R(k,\p)$ is invariant under $\sigma_0$.

\vspace{0.5truecm}
Now the problem of finding the gradients of $R$ reduces to
determine the gradients of the Lagrange resolvents. To this end we
differentiate \eqref{21} and obtain
\begin{equation}\label{35}
\left\{\begin{array}{lll}
\nonumber \mathcal{L}_2\nabla_{\bu} \mathcal{L}_1+\mathcal{L}_1
\nabla_{\bu} \mathcal{L}_2&=(2w\, ,3\, ,0)^{\top},\\\\
\nonumber \mathcal{L}_1^2\nabla_{\bu} \mathcal{L}_1+
\mathcal{L}_2^2\nabla_{\bu} \mathcal{L}_2 &=(2w^2+3v\, ,3w\,
,9)^{\top},
\end{array}\right.\end{equation}
so that
\begin{equation}\label{36}
\left\{\begin{array}{lll}
\nonumber (\mathcal{L}_1^3-\mathcal{L}_2^3)\nabla_{\bu}
\mathcal{L}_1&= \Big((2w^2+3v)\mathcal{L}_1-2w\mathcal{L}_2^2\, ,
3(w\mathcal{L}_1-\mathcal{L}_2^2)\,,9\mathcal{L}_1\Big)^{\top},
\\\\
\nonumber (\mathcal{L}_2^3-\mathcal{L}_1^3)\nabla_{\bu}
\mathcal{L}_2&= \Big((2w^2+3v)\mathcal{L}_2-2w\mathcal{L}_1^2\, ,
3(w\mathcal{L}_2-\mathcal{L}_1^2)\,,9\mathcal{L}_2\Big)^{\top}.
\end{array}\right.\end{equation}
Hence the gradients of the generic density $R$ for \eqref{31}
and \eqref{34} are
given as follows:

\vspace{0.3truecm}
For $l_0=3$, we have
\[ \begin{array}{llll}
\nabla_{\bu}R=&\displaystyle{\frac{zf_1(z^3)}{\mathcal{L}_1^3-\mathcal{L}_2^3}}
\left(\begin{array}{c}(2w^2+3v)\mathcal{L}_1-
2w\mathcal{L}_2^2\\\\
3(w \mathcal{L}_1-\mathcal{L}_2^2)\\\\
9\mathcal{L}_1\end{array}\right) \\\\
&-\displaystyle{\frac{z^2f_2(z^3)}{\mathcal{L}_1^3-\mathcal{L}_2^3}}
\left(\begin{array}{c}
(2w^2+3v)\mathcal{L}_2-
2w\mathcal{L}_1^2
 \\\\
3(w \mathcal{L}_2-\mathcal{L}_1^2)\\\\
9\mathcal{L}_2
\end{array}
\right)
+f_3(z^3)\begin{pmatrix}1 \\ 0 \\ 0 \end{pmatrix}
\end{array}
\]

\vspace{0.3truecm}
For $l_0=2$,  we get
\begin{align*}
\nabla_{\bu}R=&\frac{f_1(z^2)}{\mathcal{L}_1^3-\mathcal{L}_2^3}
\left(\begin{array}{c}(2w^2+3v)(\mathcal{L}_1-\mathcal{L}_2)
+2w(\mathcal{L}_1^2-\mathcal{L}_2^2)
 \\\\
3(w \mathcal{L}_1-\mathcal{L}_2^2)-3(w
\mathcal{L}_2-\mathcal{L}_1^2)\\\\
9(\mathcal{L}_1-\mathcal{L}_2)
\end{array}
\right)\\\\
 &+\frac{zf_2(z^2)}{\mathcal{L}_1^3-\mathcal{L}_2^3}
\left(\begin{array}{c}(2w^2+3v)(\mathcal{L}_1+\mathcal{L}_2)
-2w(\mathcal{L}_1^2+\mathcal{L}_2^2)
 \\\\
3(w \mathcal{L}_1-\mathcal{L}_2^2)+3(w
\mathcal{L}_2-\mathcal{L}_1^2)\\\\
9(\mathcal{L}_1+\mathcal{L}_2)
\end{array}
\right)+f_3(z^2)\begin{pmatrix} 1\\0\\0\end{pmatrix}.
\end{align*}
 From these expressions and \eqref{30} and \eqref{33} it follows
that the corresponding components of $\nabla_{\bu}R$ are in
$\C((k))$.

\vspace{0.5truecm}
\noindent {\bf Example 1:} The case $l_0=3$ with $(d_1,d_2,d_3)=(0,0,1)$.
Taking into account \eqref{25} and \eqref{20} it is clear that
there are two trivial equations corresponding to $w_0$ and $u_1$.
Then, we take for the potentials
$$w=1,\quad v=v_0(x,t), \quad u=k+u_0(x,t).$$
Thus, by using \eqref{31} with
\[
\quad f_1\equiv f_3\equiv 0,\quad
f_2(z^3)=\frac{27(1-\sqrt{3}i)}{4}z^3
\]
we obtain
\[
\left\{\everymath{\displaystyle}\begin{array}{rcl}
v_{0\,t}&=& \frac{5}{3}\,\left( 2 + 27\,u_0+ 9\,v_0\right)\,u_{0\,x}+\\  \\
        & & \frac{5}{18}\left( 7 + 54\,u_0 + 36\,v_0 +27\,v_0^2 \right) \,v_{0\,x},\\  \\
u_{0\,t}& =& \frac{5}{18}\,\left( -1 - 54\,u_0 +
          27\,{v_0}^2 \right) \,u_{0\,x} +\\  \\
         &  &\frac{5}{9}\,v_0\,\left( 2 + 27\,u_0 + 9\,v_0\right) \,v_{0\,x}.
\end{array}\right.
\]

It can be checked that this system corresponds to the one obtained
by setting  $M=0$, $N=3$ in \eqref{15},  and $\alpha=0$, $n=5$ in
\eqref{17}.\\  \\

\noindent {\bf Example 2:} The case $l_0=2$ with $(d_1,d_2,d_3)=(0,1,0)$,  $(l_0=2)$.
  From \eqref{25} and \eqref{20} we see that $v_{1\,t}=0$. We then take
$$w=w_0(x,t),\quad v=k+v_0(x,t), \quad u=u_0(x,t),$$
and
\[
f_1(z^2)=z^4,\quad f_2\equiv f_3\equiv0.
\]
Thus it follows
\[
\left\{\everymath{\displaystyle}\begin{array}{rcl}
w_{0\,t}& = & 4\,\left(w_0\,u_{0\,x} +v_0\,v_{0\,x}+u_0\,w_{0\,x}\right),\\  \\
v_{0\,t}& = & -2\,\left(w_0^2\,u_{0\,x} -2\,u_0\,v_{0\,x}+u_0\,w_0\,w_{0\,x}\right) +\\  \\
        &   &2v_0\,\left( 2\,u_{0\,x}-w_0\,v_{0\,x} \right),\\  \\
u_{0\,t}& = &-2\,\left( v_0\,w_0\,u_{0\,x} + u_0\,\left(
-2\,u_{0\,x}+w_0\,v_{0\,x} +
              v_0\,w_{0\,x} \right)  \right).
\end{array}\right.
\]
It turns out that this system can also be found among the Whitham
deformations, by setting $M=1$, $N=2$ in \eqref{15}, and
$\alpha=0$,
$n=4$ in \eqref{17}.\\  \\

\noindent
{\bf Example 3:} The case $l_0=2$ with  $(d_1,d_2,d_3)=(1,0,0)$.
  From \eqref{20} and \eqref{25} it is easy to see that
$\left(\displaystyle\frac{u_0}{w_1}\right)_t=0$. If we choose
\[
\left\{\begin{array}{ll}
w(k,x,t)=w_1(x,t)k+w_0(x,t),\quad v(k,x,t)=v_0(x,t),\\  \\
 u(k,x,t)=w_1(x,t),
\end{array}\right.
\]
and set
$$f_1(z^2)=z^4,\quad f_2\equiv f_3\equiv 0,$$
in  \eqref{34}, then the following system arises
\[\left\{\everymath{\displaystyle}\begin{array}{lcl}
w_{1\,t} & = & 2\,w_1^{-2}(w_1\,w_{0\,x}\,-\,w_0\,w_{1\,x}),\\  \\
w_{0\,t} & = & 2w_1^{-3}\left(w_1\,\left(v_{0\,x}
+w_0\,w_{0\,x}\right)  -
       \left( 2\,v_0 + w_0^2 \right) \,w_{1\,x}\right),\\  \\
v_{0\,t} & = & w_1^{-3}
\left(-4\,w_1\,w_{1\,x}\,+\,2\,v_0\,\left(w_1\,w_{0\,x}\, -\,
          w_0\,w_{1\,x}\right)\right).
\end{array}\right.\]
This is one of the flows in the dispersionless Dym hierarchy
corresponding to the curve, $w_1k=p-w_0-v_0p^{-1}-w_1p^{-2}$.
Also note that the linear flow, i.e. $w_{1t}=cw_{1x}$ etc with $c=$constant,
can be obtained by the choice $f_2\propto z^{-2}$ with $ f_1=f_3=0$.\\ \\

\noindent
{\bf Example 4:} The case $l_0=1$ with $(d_1,d_2,d_3)=(1,0,1)$.
  From \eqref{20} and \eqref{25} one finds that
$\left(\displaystyle\frac{u_1}{w_1}\right)_t=0$. By setting
\[\left\{\begin{array}{ll}
w(k,x,t)=w_1(x,t)k+w_0(x,t),\quad v(k,x,t)=v_0(x,t),\\  \\
u(k,x,t)=w_1(x,t)k+u_0(x,t),
\end{array}\right.\]
and
$$
R=\frac{2(1+\sqrt{3}i)}{\sqrt{3}}k\mathcal{L}_1,$$ we obtain
\[\left\{\everymath{\displaystyle}\begin{array}{lcl}
 w_{1\,t} & = &  u_{0\,x}\,+\,w_{0\,x},\\  \\
 w_{0\,t} & = & w_1^{-2}\Big( {w_1}\,\left(  v_{0\,x}\, +\,u_0\,w_{0\,x} \right)  -
     \left( 3 +  v_0 \right) \,w_{1\,x} \,-\, w_0^2\,w_{1\,x} \\  \\
          &   &+ w_0\,\left(w_1\,
         \left(u_{0\,x} \,+\, 2\,w_{0\,x} \right)\,  -\,
        u_0\, w_{1\,x} \right)  \Big),\\  \\
v_{0\,t} & = & w_1^{-2}\Big(w_1\,\left( 2\,\left( 2 +v_0\right) \,
u_{0\,x} +
         u_0\,v_{0\,x}\, + \,w_0\,v_{0\,x} + 2\,v_0\,w_{0\,x} \right) \\  \\
         &    & - 2\,\left(u_0\,\left( 3 +v_0\right) \,-\,\left(1 - v_0 \right) \,w_0 \right) \, w_{1\,x}\Big),\\  \\
u_{0\,t} & = & w_1^{-2}\Big(-w_0\,w_1\,u_{0\,x} -
3\,u_0^2\,w_{1\,x} +
       v_0\,\left(w_1\,v_{0\,x} + \left(1 - v_0\right) \, w_{1\,x} \right) \\  \\
         &    &+  u_0\,\left(w_1\,\left( 4\, u_{0\,x} + w_{0\,x} \right)  +
         w_0\,w_{1\,x} \right) \Big).
\end{array}\right.\]
We also note that the linear flow is obtained by choosing $R\propto \mathcal{L}_1$,
and the higher flows in  the hierarchy can be obtained by $R\propto k^n\mathcal{L}_1$.\\ \\

\noindent
{\bf Example 5:} The case $l_0=1$ with  $(d_1,d_2,d_3)=(1,1,0)$.
  From \eqref{20} and \eqref{25} we deduce that
$\left(\displaystyle\frac{v_1}{w_1}\right)_t=0$. Now we take
\[\left\{\begin{array}{ll}
w(k,x,t)=w_1(x,t)k+w_0(x,t),\quad v(k,x,t)=w_1(x,t)k+v_0(x,t),\\  \\
u(k,x,t)=u_0(x,t)
\end{array}\right.\]
and set
$$
R=\frac{\sqrt{3}+i}{2\sqrt{3}}k\mathcal{L}_2\,.$$
Then the following
system is obtained
\[\left\{\everymath{\displaystyle}\begin{array}{lcl}
w_{1\,t} & = &2\, u_{0\,x}-v_{0\,x},\\  \\
w_{0\,t} & = & w_1^{-2}\Big( w_1\,\left( \left( 3 + 2\, w_0
\right) \, u_{0\,x} -
        \left( 2 + w_0 \right) \, v_{0\,x} +
        \left( 2\, u_0- v_0 \right) \,w_{0\,x} \right)\\  \\
         &    &+ \left( v_0\,\left( 2 +  w_0 \right)  -
         u_0\,\left( 3 + 2\, w_0 \right)  \right) \,w_{1\,x}\Big),\\  \\
v_{0\,t} & = & w_1^{-2} \Big(w_1\,\left( \left( -2 + 4\,v_0 -
2\,w_0 \right) \,u_{0\,x}
        + \left( 2\, u_0 - 3\, v_0+ w_0 \right) \, v_{0\,x} \right) \\  \\
         &    &+  \left(v_0\,\left( 2\,v_0 -w_0 \right)  +
         u_0\,\left( 3 - 4\,v_0 + 2\,w_0\right)  \right) \,w_{1\,x}\Big),\\  \\
u_{0\,t} & = & w_1^{-2} \Big(\left(-2\, v_0 +w_0 \right) \,
       w_1\,u_{0\,x}- 6\,u_0^2\,w_{1\,x} \\   \\
         &   &+u_0\,\left(w_1\,\left( 8\, u_{0\,x} - 3\,v_{0\,x} +
            w_{0\,x} \right)  +
        2\,\left( 2\, v_0 -  w_0 \right) \, w_{1\,x}        \right) \Big).
\end{array}\right.\]

\subsection{Hamiltonian structures}

The general structure of integrable deformations \eqref{7} does
not exhibit a direct hamiltonian form. However, the analysis of
particular cases reveals the presence of certain Hamiltonian
structures. We look for a Hamiltonian operator $J$ such that for
certain appropriate densities $R$ it verifies
\begin{equation}\label{37}
J_0\Big(T\,\nabla_{\bu} R\Big)_{+}=J\Big(\nabla_{\bu} R\Big)_{+},
\end{equation}
where
\[
T:=\left(
\begin{array}{ccc}
1&-w&-v\\
0&1&-w\\
0&0&1
\end{array}
\right).
\]
Thus, if \eqref{37} holds then the flows  \eqref{7} can be written
in the pre-Hamiltonian form
\begin{equation}\label{38}
\partial_t \bu = J\Big(\nabla_{\bu} R\Big)_{+}.
\end{equation}

To achieve our  aim we require a $k$-independent operator $T_0$
verifying
\begin{equation}
T\,\nabla_{\bu} R=T_0\,\nabla_{\bu} R,
\end{equation}
so that $J:=J_0\cdot T_0$ is a Hamiltonian operator.

Let us consider first the case $l_0=3$. It involves two classes of
cubic curves:

\vspace{0.3truecm}
For the case with $(d_1,d_2,d_3)={(0,0,1)}$,
the potentials  are of the form
\[
w=w_0(x),\quad v(x)=v_0(x),\quad u=u_0(x)+k\, u_1(x).
\]
The matrix $T$ is $k$-independent so that by setting $J=J_0\cdot
T$ we find the Hamiltonian operator
\begin{equation}\label{40}
J=
\left(
\begin{array}{ccc}
3\partial_x & -2\partial_x\cdot w&-\partial_x\cdot v \\\\
-2w\,\partial_x& 2w\partial_x\cdot w+2v\partial_x+v_x
&(3u+vw)\partial_x+2u_x+wv_x\\\\
-v\partial_x&(3u+vw)\partial_x+vw_x+u_x&v\partial_x \cdot v-
2uw\partial_x-(u w)_x
\end{array}
\right).
\end{equation}
It represents the dispersionless limit of the Hamiltonian structure
of the Boussinesq hierarchy.

\vspace{0.3truecm}
For the case with ${(0,1,2)}$
the potentials now are
\[ \left\{\begin{array}{lll}
w=w_0(x),\quad v(x)=v_0(x)+k\, v_1(x),\\
u=u_0(x)+ k\, u_1(x)+k^2\, u_2(x).
\end{array}\right.\]
  From \eqref{36} one deduces
\begin{equation}\label{41}
\left\{\begin{array}{ll}
T\,\nabla_{\bu} \mathcal{L}_i=T_0\,\nabla_{\bu}\mathcal{L}_i,\quad
i=1,2,\\
T\,\nabla_{\bu} \mathcal{L}_3=\mathcal{L}_3,
\end{array}\right.
\end{equation}
where $T_0$ is the $k$-independent matrix
\begin{equation}\label{42}
T_0=\left(
\begin{array}{ccc}
-2&w&0\\
0&1&-w\\
0&0&1
\end{array}
\right).
\end{equation}
Moreover $J:=J_0\cdot T_0$ takes the Hamiltonian form
\begin{equation}\label{43}
J=
\left(
\begin{array}{ccc}
-6\partial_x & 4\partial_x\cdot w& 2\partial_x\cdot v \\\\
4w\,\partial_x& -2w\partial_x\cdot w+2v\partial_x+v_x
&(3u-vw)\partial_x+2u_x-wv_x\\\\
2v\partial_x&(3u-vw)\partial_x-vw_x+u_x& -2uw\partial_x-(u w)_x
\end{array}
\right).
\end{equation}
Thus by setting
\[
R=zf_1(z^3)\mathcal{L}_1+z^2f_2(z^3)\mathcal{L}_2,
\]
equation \eqref{7} reduces to the form \eqref{38}.

For the remaining cases of $l_0=2$ and $l_0=1$, the situation is as follows:
\begin{enumerate}
\item For the sets of degrees $(0,1,0)$ and $(0,1,1)$ for $l_0=2$, the
identities \eqref{41} with the same operator \eqref{42} hold, so
that by setting
\[
R=f_1(z^2)\Big( \mathcal{L}_1+\mathcal{L}_2\Big)+zf_2(z^2)\Big( \mathcal{L}_1- \mathcal{L}_2\Big),
\]
equation \eqref{7} reduces to the form \eqref{38} with the
Hamiltonian operator \eqref{43}.
\item For the sets of degrees (two cases of $l_0=2$ and all the cases of $l_0=1$),
\begin{align}\label{34}
\nonumber
&(1,0,0),\, (1,0,1),\,(1,1,0),\,(1,1,1)\\
\nonumber &(1,1,2),\,(1,2,1),\,(1,2,2),\, (1,2,3),
\end{align}
there is no $k$-independent operator $T_0$ satisfying \eqref{40}
for $\nabla_{\bu}\mathcal{L}_i,\, (i=1,2)$.
\end{enumerate}

\subsection{Deformations of cubic curves with $w=0$}

Deformations of cubic curves of the form
\begin{equation}\label{44}
p^3-v\,p-u=0,
\end{equation}
cannot be obtained simply by setting $w=0$ in the above analysis.
Indeed, as it is clear from the expression \eqref{25} for $J_0$,
the constraint $w=0$ does not constitutes a reduction of the flows
\eqref{7}. Therefore, we have to apply our deformation scheme  to
\eqref{44} directly.

In terms of the branches $p_i$ the condition $w=0$ reads
\[
p_1+p_2+p_3=0,
\]
which is preserved by deformations
\begin{equation}\label{45}
\partial_t p_i=\partial_x(a_1+a_2\,p_i+a_3\,p_i^2)
\end{equation}
satisfying
\begin{equation}\label{46}
3a_1=-(p_1^2+p_2^2+p_3^2)\,a_3.
\end{equation}
By expressing the potentials as functions of the branches $p_1$
and $p_2$
\begin{equation}\label{47}
v=p_1^2+p_2^2+p_1p_2,\quad u=-(p_1^2p_2+p_1p_2^2),
\end{equation}
and using \eqref{45} and \eqref{46}, we obtain
\begin{equation}\label{48}
\partial_t \bu = \mathcal{J}_0\,\ba,\quad
\bu:=(v\quad u)^{\top},\quad
 \ba:=(a_1\quad a_2)^{\top},
\end{equation}
where
\begin{align}\label{49}
\nonumber \mathcal{J}_0&= \left(\begin{array}{cc}
2p_1+p_2&2p_2+p_1\\
-2p_1p_2-p_2^2&-2p_1p_2-p_1^2
\end{array}\right)\partial_x
\left(\begin{array}{cc}
p_1&\frac{1}{3}p_1^2-\frac{2}{3}(p_2^2+p_1p_2)\\
p_2&\frac{1}{3}p_2^2-\frac{2}{3}(p_1^2+p_1p_2)
\end{array}\right)\\\\
\nonumber &=\left(\begin{array}{cc}
2v\partial_x+v_x&3u\partial_x+2u_x\\
3u\partial_x+u_x&\frac{1}{3}(2v^2\partial_x+2vv_x)
\end{array}\right).
\end{align}

According to our strategy for finding consistent  deformations, we use Lenard
type relations
\[
\mathcal{J}_0\bR=0,\quad \bR:=(r_1\quad r_2)^\top,\;\;
r_i\in\C((k)),
\]
to generate systems of the form
\begin{equation}\label{50}
\bu_t=\mathcal{J}_0\ba,\quad \ba:=\bR_+.
\end{equation}
Here $(\,\cdot\,)_+$ and $(\,\cdot\,)_-$ indicate the parts of
non-negative and negative powers in $k$, respectively. Now from the identity
\[
\mathcal{J}_0\ba =\mathcal{J}_0\bR_+=-\mathcal{J}_0\bR_-,
\]
it is clear that  a sufficient condition for the consistency of
\eqref{50} is that the degrees $d_2$ and $d_3$ of $v$ and $u$ as
polynomials of $k$ satisfy
\[
d_3\leq d_2+1,\quad 2d_2\leq d_3+1.
\]
Hence only four nontrivial cases arise for $(d_2,d_3)$
\begin{equation}\label{51}
(0,1),\quad (1,1),\quad (1,2),\quad (2,3).
\end{equation}
We notice that they represent the dispersionless versions of the
standard Boussinesq hierarchy and all three hidden hierarchies
found by Antonowicz, Fordy and Liu for the third-order spectral
problem \cite{16}.

Solutions of the Lenard relation can be generated by noticing that
the operator $\mathcal{J}_0$ admits the factorization
\begin{equation}\label{52}
\mathcal{J}_0=U^{\top}\cdot\frac{1}{3} \left(
\begin{array}{cc}
2&-1\\
-1&2
\end{array}\right)\partial_x\,\cdot U,
\end{equation}
where
\begin{equation}\label{53}
U:= \left(\begin{array}{cc}
2p_1+p_2&-2p_1p_2-p_2^2\\
2p_2+p_1&-2p_1p_2-p_1^2
\end{array}\right)
=\left(\begin{array}{cc}
\frac{\partial v}{\partial p_1}&\frac{\partial u}{\partial p_1}\\
\frac{\partial v}{\partial p_2}&\frac{\partial u}{\partial p_2}
\end{array}\right).
\end{equation}
This shows two things:
\begin{description}
\item[i)]  $\mathcal{J}_0$ is a Hamiltonian operator.
\item[ii)] The gradients $\nabla_{\bu} p_i$ of the branches $p_1$ and $p_2$
solve the Lenard relations.
\end{description}
Thus our candidates to deformations
are the equations of the form
\begin{equation}\label{54}
\partial_t \bu = \mathcal{J}_0\Big(\nabla_{\bu} R\Big)_{+},\quad R(z,\p)=f_1(z)\,p_1+f_2(z)\,p_2,
\end{equation}

At this point one applies the same strategy as that used for the
curves \eqref{3} in subsection 3.1. We first determine the  Newton
exponents of the four cases \eqref{51} which turn to be given by
\medskip

 \hfil\vbox{\hbox{\vbox{\offinterlineskip
 \halign{&\vrule#&\strut\quad#\hfil\quad&\vrule#&
 \quad#\hfil\quad&\vrule#&
 \quad#\hfil\quad\cr
 % '\quad#\hfil\quad' in the lines above
 % gives left aligned entries. Substitute
 % `\quad\hfil#\hfil\quad' or '\quad\hfil#\quad'
 % for centered or right aligned.
 \noalign{\hrule}
 height2pt&\omit&&\omit&&\omit&&\omit&\cr
 &$l_0$&&\hspace{0.4truecm}3&&\hspace{0.4truecm}2&&\hspace{0.4truecm}1&\cr
 height2pt&\omit&&\omit&&\omit&&\omit&\cr
 \noalign{\hrule}
 height2pt&\omit&&\omit&&\omit&&\omit&\cr
   &  && (0,1) && (1,1) && (2,3) &\cr
   &  &&(1,2) &&  &&  &\cr
 height2pt&\omit&&\omit&&\omit&&\omit&\cr
 \noalign{\hrule}}}}}\hfill

\medskip

\noindent
Then,  with the help of Lagrange resolvents, we characterize
the functions $R(z,\p)$ verifying
$\nabla_{\bu}R\in \C((k))$ with $ k=z^{l_0}$. In summary, one finds

\vspace{0.3truecm}
For the case ${l_0=3}$,
\begin{equation}\label{55}
R=zf_1(z^3)\,\mathcal{L}_1+z^2f_2(z^3)\, \mathcal{L}_2,\quad
k=z^3.
\end{equation}

For the case ${l_0=2}$,
\begin{equation}\label{56}
R=f_1(z^2)\Big( \mathcal{L}_1+\mathcal{L}_2\Big)
 +zf_2(z^2)\Big( \mathcal{L}_1- \mathcal{L}_2\Big),\quad k=z^2.
\end{equation}

For the case ${l_0=1}$, we have $z=k$, so that any function $R(k,\p)=
f_1(k)\mathcal{L}_1+f_2(k)\mathcal{L}_2$ is appropriate.

\vspace{0.5truecm}
\noindent
{\bf Example 1:} The case $l_0=3$ with
$(d_2,d_3)=(1,2)$. From \eqref{48} and \eqref{49} we have that $u_{2\,t}=0$. Then if
one takes
$$u(k,x,t)=k^2+u_1(x,t)k+u_0(x,t),\qquad
v(k,x,t)=v_1(x,t)k+v_0(x,t),$$ and sets
$$f_1(z^3)=\frac{1}{2}(1+i\sqrt{3})z^3,\qquad f_2\equiv0,$$
in \eqref{55}, one gets
\[\left\{\everymath{\displaystyle}\begin{array}{lcl}
v_{1\,t} & = & -2\,u_{1\,x} +\frac{5}{9}\,v_1^2\,v_{1\,x},\\  \\
v_{0\,t} & = &\frac{1}{9}\left(-18\,u_{0\,x} + \, v_1^2\,v_{0\,x} +4 \,v_0\,v_1\,v_{1\,x}\right),\\  \\
u_{1\,t} & = & \frac{1}{9}\left(v_1^2\,u_{1\,x} -6\,v_0\,v_{1\,x} - 6\,v_1\, v_{0\,x} + 6\,v_1\,u_1\,v_{1\,x}\right),
                     \\  \\
u_{0\,t} & = & \frac{1}{9}\left(v_1^2\,u_{0\,x} - 6\,v_0\,v_{0\,x}
+ 6\,u_0\,v_1\,v_{1\,x}\right),
\end{array}\right.\]
i.e., the dispersionless version of the {\em coupled Boussinesq
system} (3.20b) in \cite{16}.

\vspace{0.5truecm}
\noindent
{\bf Example 2:} The case $l_0=2$ with
$(d_2,d_3)=(1,1)$.
Now, one can
see that $v_{1\,t}=0$. By setting
$$u(k,x,t)=u_1(x,t)k+u_0(x,t),\qquad
v(k,x,t)=-k+v_0(x,t),$$ and
$$f_1(z^2)=-z^2,\qquad f_2\equiv0,$$
in \eqref{56}, we find the system,
\[\left\{\everymath{\displaystyle}\begin{array}{lcl}
v_{0\,t}& = & -2\,u_{0\,x} - 2\,v_0\,u_{1\,x}-u_1\,v_{0\,x},\\  \\
u_{1\,t}& = & -4\,u_1\,u_{1\,x} + \frac{2}{3}\,v_{0\,x},\\  \\
u_{0\,t}& = & -u_1\,u_{0\,x} -3\,u_0\,u_{1\,x}
-\frac{2}{3}v_0\,v_{0\,x}\,.
\end{array}\right.\]
This is the dispersionless version of the system (4.13) in
\cite{16}.

\subsection{Whitham deformations of cubic curves}

There are four types of cubic curves of the form \eqref{15} given
by the equations
\begin{center}
 \hfil\vbox{\hbox{\vbox{\offinterlineskip
 \halign{&\vrule#&\strut\quad#\hfil\quad&\vrule#&
 \quad#\hfil\quad&\vrule#&
 \quad#\hfil\quad\cr
 % '\quad#\hfil\quad' in the lines above
 % gives left aligned entries. Substitute
 % `\quad\hfil#\hfil\quad' or '\quad\hfil#\quad'
 % for centered or right aligned.
 \noalign{\hrule}
 height2pt&\omit&&\omit&&\omit&&\omit&\cr
 &$M$&&\hspace{1.5truecm}0&&\hspace{1.7truecm}1&&\hspace{1.7truecm}2&\cr
 height2pt&\omit&&\omit&&\omit&&\omit&\cr
 \noalign{\hrule}
 height2pt&\omit&&\omit&&\omit&&\omit&\cr
   &  && $k=p^3+v_1\,p+v_0$ && $k=p^2+v_0+\frac{v_1}{p-w_1}$ && $k=p+\frac{v_{1,1}}{p-w_1}+\frac{v_{2,1}}{p-w_2}$ &\cr
   &  && && $k=p+\frac{v_1}{p-w_1}+
\frac{v_2}{(p-w_1)^2}$  &&  &\cr
 height2pt&\omit&&\omit&&\omit&&\omit&\cr
 \noalign{\hrule}}}}}\hfill
\end{center}
Note here that the Newton exponent $l_0$ is given by $l_0=3-M$. Also
in \cite{8}, two cases in $M=1$ are shown to be conformally equivalent,
i.e. $p=\infty \leftrightarrow p=w_1$.

\vspace{0.3truecm}

For $M=0$ the Whitham deformations are reductions of our flows
with $w\equiv 0$. But, in general, other Whitham deformations are
not of that form. To illustrate this point let us take the class
with $M=1$ and $N=2$. The corresponding Newton exponent is $l_0=2$
and there the branches of $\p$ have the following asymptotic
behaviour as $z\rightarrow\infty$
\[\left\{\begin{array}{lll}
p_1(z)=z+\mathcal{O}\Big(\frac{1}{z}\Big),\\
p_2(z)=p_1(-z)=-z+\mathcal{O}\Big(\frac{1}{z}\Big),\\
p_3(z)=w_1+\mathcal{O}\Big(\frac{1}{z}\Big).
\end{array}\right.\]
Let us consider now the Whitham flows \eqref{17} associated with
the puncture at $p=\infty$
\[
\Q_{0,n}=(z^n)_{\oplus}(\p).
\]
In terms of the potentials $\bu=(w,v,u)^{\top}$ they read
\begin{equation}\label{57}
\partial_t \bu = J_0{\bf a},
\end{equation}
where
\[
{\bf a}=\left(V^{-1}
\left(\begin{array}{c}
(z^n)_{\oplus}(p_1)\\
(z^n)_{\oplus}(p_2)\\
(z^n)_{\oplus}(p_3)
\end{array}
\right) \right)_+.
\]
One easily sees that all matrix elements of $V^{-1}$ are of order
$\mathcal{O}\Big(\frac{1}{z}\Big)$ with the exception of
\[
\Big(V^{-1}\Big)_{13}=1+\mathcal{O}\Big(\frac{1}{z}\Big).
\]
On the other hand, we have
\[\left\{\begin{array}{lll}
(z^n)_{\oplus}(p_1)=&z^n+\mathcal{O}\Big(\frac{1}{z}\Big),\\
(z^n)_{\oplus}(p_2)=&(-z)^n+\mathcal{O}\Big(\frac{1}{z}\Big),\\
(z^n)_{\oplus}(p_3)=&(z^n)_{\oplus}(w_1)+\mathcal{O}\Big(\frac{1}{z}\Big).
\end{array}\right.
\]
Therefore one gets
\[
{\bf a}=\left(z^n V^{-1}
\begin{pmatrix}
1\\
(-1)^n\\
0
\end{pmatrix}
\right)_++(z^n)_{\oplus}(w_1)\, {\bf e}_3,
\]
where ${\bf e}_3=(0,0,1)^{\top}$, so that equation \eqref{57} becomes
\begin{equation}\label{58}
\partial_t \bu = J_0\Big(T\,\nabla_{\bu}[z^n(p_1+(-1)^n p_2)]\Big)_{+}+
J_0\Big((z^n)_{\oplus}(w_1)\,{\bf e}_3\Big).
\end{equation}
Similar expressions can be obtained for the deformations generated
by the Whitham flows \eqref{17} for $\alpha=1$ and $n\geq 1$.

\vspace{0.5truecm}
\noindent
{\bf Acknowledgements}

\vspace{0.3truecm}
\noindent

L. Mart\'{\i}nez Alonso wishes to thank the members of the Physics
Department of Lecce  University  for their warm hospitality.

\end{document}